\documentclass{article}

\usepackage{algorithm, algpseudocode}
\usepackage{amsfonts, amsmath, amsopn, amsthm, epsfig}
  \numberwithin{equation}{section}
\usepackage{graphicx, epstopdf}
\usepackage{hyperref}
\usepackage{subfigure}
\usepackage{appendix}
\usepackage{authblk}

\usepackage{calc}
\usepackage[top=2.5cm,bottom=2.3cm,inner=2.3cm,outer=2.3cm,bindingoffset=0.5cm,footskip=0.55cm]{geometry}
\theoremstyle{remark}

\newenvironment{lemma*}[2][Lemma]{\par\bgroup{\bfseries #1\ #2. }\it\ignorespaces}{\egroup}

\title{Hodge and Podge: Hybrid Supervised Sound Event Detection with Multi-Hot MixMatch and Composition Consistence Training
}

\author[]{Ziqiang Shi\thanks{Corresponding author: shiziqiang@cn.fujitsu.com; shiziqiang7@gmail.com}}
\author[]{Liu Liu}
\author[]{Huibin Lin}
\author[]{Rujie Liu}

\affil[]{Fujitsu Research and Development Center, Beijing, China}

\newcommand{\BALD}{\begin{aligned}}
\newcommand{\EALD}{\end{aligned}}
\newcommand{\BALDS}{\begin{aligned*}}
\newcommand{\EALDS}{\end{aligned*}}
\newcommand{\BCAS}{\begin{cases}}
\newcommand{\ECAS}{\end{cases}}
\newcommand{\BEAS}{\begin{eqnarray*}}
\newcommand{\EEAS}{\end{eqnarray*}}
\newcommand{\BEQ}{\begin{equation}}
\newcommand{\EEQ}{\end{equation}}
\newcommand{\BIT}{\begin{itemize}}
\newcommand{\EIT}{\end{itemize}}
\newcommand{\BMAT}{\begin{bmatrix}}
\newcommand{\EMAT}{\end{bmatrix}}
\newcommand{\BNUM}{\begin{enumerate}}
\newcommand{\ENUM}{\end{enumerate}}

\newcommand{\BA}{\begin{array}}
\newcommand{\EA}{\end{array}}

\begin{document}

\maketitle

\begin{abstract}
In this paper, we propose a method called Hodge and Podge for sound event detection. We demonstrate Hodge and Podge on the dataset of Detection and Classification of Acoustic Scenes and Events (DCASE) 2019 Challenge Task 4. This task aims to predict the presence or absence and the onset and offset times of sound events in home environments. Sound event detection is challenging due to the lack of large scale real strongly labeled data. Recently deep semi-supervised learning (SSL) has proven to be effective in modeling with weakly labeled  and unlabeled data. This work explores how to extend deep SSL to result in a new, state-of-the-art sound event detection method called Hodge and Podge. With convolutional recurrent neural networks (CRNN) as the backbone network, first, a multi-scale squeeze-excitation mechanism is introduced and added to generate a pyramid squeeze-excitation CRNN. The pyramid squeeze-excitation layer can  pay attention to the issue that different sound events have different durations, and to adaptively recalibrate channel-wise spectrogram responses. Further, in order to remedy the lack of real strongly labeled data problem, we propose multi-hot MixMatch and composition consistency training with temporal-frequency augmentation. Our experiments with the public DCASE2019 
challenge task 4 validation data resulted in an event-based F-score of 43.4\%, and is about absolutely 1.6\% better than state-of-the-art methods in the challenge. While the F-score of the official baseline is 25.8\%.
\end{abstract}

\section{Introduction}
\label{sec:introduction}

Recently, sound event detection (SED) has become more and more popular in the field of acoustic signal processing, since it can be widely used in everyday life.  By leveraging large corpora of strongly annotated sound data, where the onset and offset times of sound events have been annotated, that allow state-of-the-art models such as deep neural networks (DNN) can learn the characteristics of sound events to tackle the problem of SED. However, acquiring such large amounts of strongly labeled data is practically infeasible, since annotating the onset and offset times of sound events take more time than annotating audio clips for classification. 
%For example, SED can be used to detect “smoke alarms” and “sirens” in the office, to monitor “baby cry” sound at home, and to detect “scream” and  “gunshot” in public area~\cite{ellis2001detecting,harma2005automatic,saraswathy2012automatic}. One important application of SED in the smart home is home activity detection and monitoring. 

Obviously, if there is no real strongly labeled data, then the data can use is only weakly labeled data, maybe synthesized data, and unlabeled data. In SED, weak labels refer to ground truth labels that where only the presence of the sound events is labeled, with no temporal information on the onset nor offset of audio events. In addition to the weakly labeled data, we can also synthesize sound data to simulate the target application environment. Manually verified foreground events are embedded  in the  background texture, which is similar to the target environment. Then the distribution of sound events per class, the number of sound events per clip and the sound event class co-occurrence can be designed to be similar to the real recordings. At the same time, a large amount of unlabeled audio data can be recorded in the target environment. These unlabeled in-domain data may help us to augment supervised training methods, for example by semi-supervised learning method~\cite{zhu2005semi,chapelle2009semi}. This is the situation that we are concerned in this paper, and also is precisely the challenge proposed by the Detection and Classification of Acoustic Scenes and Events (DCASE) 2019  task 4~\cite{turpault2019sound}. 

%To contribute to the SED task, the DCASE challenge has been organized for 5 times of 7 years since 2013~\cite{mesaros2018detection,mesaros2017dcase,serizel2018large,turpault2019sound}. DCASE is a series of challenges aimed at developing sound classification and detection systems~\cite{mesaros2018detection,mesaros2017dcase,serizel2018large,turpault2019sound}.This year, the DCASE 2019 task 4 challenge focused on large-scale detection of sound events in  domestic environments using real data either weakly labeled or unlabeled and synthetic data that is strongly labeled (with timestamps)~\cite{turpault2019sound}. 

This task is the follow-up to DCASE 2018 task 4, which aims at exploring the possibility to exploit a large amount of unbalanced and unlabeled training data together with a small weakly annotated training set to improve system performance. The difference is that there is an additional training set with strongly annotated synthetic data provided in this year's task 4. Furthermore, a baseline system that performs the task is provided in the DCASE 2019 challenge~\cite{tarvainen2017mean,jiakai2018mean}, and there have been a variety of methods proposed to solve this problem~\cite{serizel2018large,jiakai2018mean,kong2018dcase}.

This paper is based on the `Mean Teacher' approach~\cite{tarvainen2017mean,jiakai2018mean}, which has achieved the best result in DCASE 2018 task 4~\cite{serizel2018large}, and also based on HODGEPODGE~\cite{shi2019hodgepodge}, which has ranked third in DCASE 2019 task 4. We propose to introduce a  multi-scale squeeze-excitation mechanism into the standard convolutional recurrent neural network (CRNN) as the backbone network for both student and teacher in `Mean Teacher' framework. In order to make full use of a small amount 
of weakly labeled and synthetic data, different temporal-frequency augmentation methods are utilized to supplement supervised training. In parallel, we introduce composition consistency training and multi-hot MixMatch training to encourage the model to produce the same mixed output distribution when its inputs are perturbed and interpolated, to generalize well and avoid overfitting on the labeled data, and to output confident predictions on unlabeled data.

%The rest of this paper is organized as follows: In Section~\ref{sec:sed}, we introduce `Mean Teacher' and Hybrid supervised sound event detection. Section~\ref{sec:method} introduces details of the proposed Hodge and Podge The experiment settings and results are displayed and discussed in Section~\ref{sec:results}. We conclude this paper in Section~\ref{sec:conc}.

\section{Hybrid Supervised Sound Event Detection}
\label{sec:sed}

Compared to the strongly or weakly supervised SED task, where strongly labeled onset and offset annotations or clip-level labels for the training set are given, the hybrid supervised SED task contains clip-level labels for weakly labeled data, strong labels for synthetic data, and unlabeled in-domain data. 

\begin{figure}[th]
\centering
\includegraphics[width=0.5\textwidth]{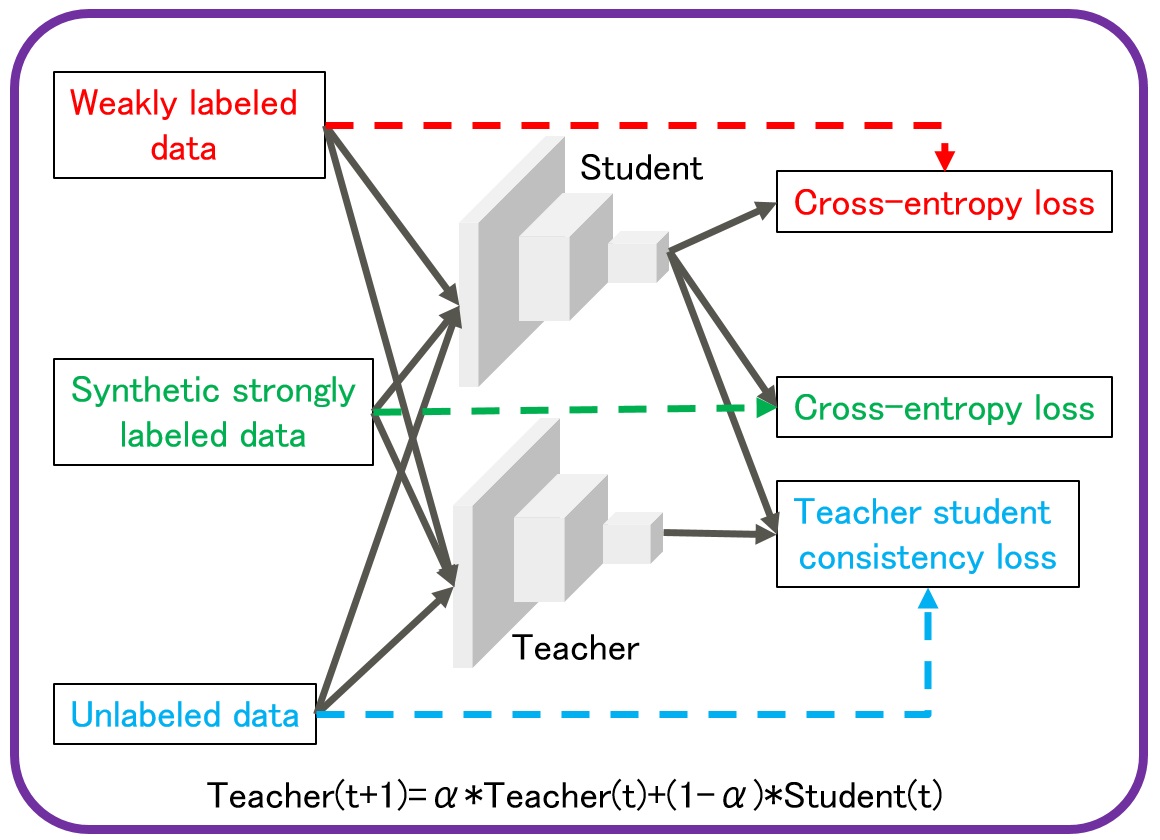}
\caption{`Mean Teacher' framework employed and adapted in Hodge and Podge.}
\label{fig:mean_teacher}
\end{figure}

In order to deal with this ill-posed problem,~\cite{jiakai2018mean} introduces the `Mean Teacher' approach~\cite{tarvainen2017mean} to do a similar task. The difference in situation is that there is no out-of-domain unlabeled audio here, but instead, synthetic strongly labeled data is easily obtained and provided. `Mean Teacher' employed and adapted in Hodge and Podge is a combination of two models: the student model and the teacher model as shown in Fig.~\ref{fig:mean_teacher}. At each training step, the student model is trained on synthetic and weakly labeled data with binary cross-entropy (CE) classification cost. While the teacher model is an exponential moving average of the student models. The student model is the final model and the teacher model is designed to help the student model by a consistency mean-squared error cost on frame-level and clip-level predictions of unlabeled audio clips. That means a good student should output the same class distributions as the teacher for the same unlabeled example even after it has been perturbed by augmentation which will be introduced in Section~\ref{ssec:preprocessing}.

%The goal of `Mean Teacher' is to minimize
%\begin{equation}
%L=L_w+L_s+w(t)(L_{cs}+L_{cw}) \nonumber
%\end{equation}
%where $L_w$ is the usual cross-entropy (CE) classification loss on originally weakly labeled data with only multi-hot weak labels, $L_s$ is the CE loss on synthetic strongly labeled data with only  frame-level multi-hot strong labels, $L_{cw}$ and $L_{cs}$ are the teacher-student consistency regularization loss on unlabeled data with predicted weak and strong multi-hot labels respectively, and $w(t)$ is the balance of classification CE loss and the consistency loss.

\section{Hodge and Podge}
\label{sec:method}

Herein, in the following sections, we will describe the details of Hodge and Podge, including feature extraction, pyramid squeeze-excitation CRNN, multi-hot MixMatch training, and composition consistency training.

\subsection{Feature Extraction and Augmentation}
\label{ssec:preprocessing}

Our proposal Hodge and Podge are developed and evaluated on the dataset provided by the DCASE 2019 challenge task 4~\cite{turpault2019sound}. The dataset for task 4 is composed of 10-sec audio clips recorded in a domestic environment or synthesized to simulate a domestic environment. No preprocessing step was applied in the presented frameworks. The acoustic features for the 44.1kHz original data used in this system consist of 128-dimensional log Mel-band energy extracted in Hanning windows of size 2048 with 431 points overlap. Thus the maximum number of frames is 1024. The input to the network is fixed to be a 10-second audio clip. If the input audio is less than 10 seconds, it is padded to 10 seconds; otherwise, it is truncated to 10 seconds.

In order to prevent the system from overfitting on the small amount of development data, we added random white noise (before log operation) to the Mel spectrogram  in each mini-batch during training. We also propose to introduce data augmentation by temporal-frequency shift. The temporal shift augmentation is a random shift of the signal by rolling the signal along the time axis. The frequency shift augmentation is a random roll in the range +-5\% around the frequency axis in the Mel domain. A wrap-around both temporal-frequency shifts to preserve all information. Here +-5\% wrap-around in frequency does not affect the sound much physically or perceptually, but can generate a lot of augmented data. One thing to note is that the frame-level labels of strongly annotated synthetic data also have to be shifted accordingly over the temporal shift.

\subsection{Pyramid Squeeze-Excitation CRNN}
\label{ssec:architecture}

\begin{figure}[th]
\centering
\includegraphics[width=0.4\textwidth]{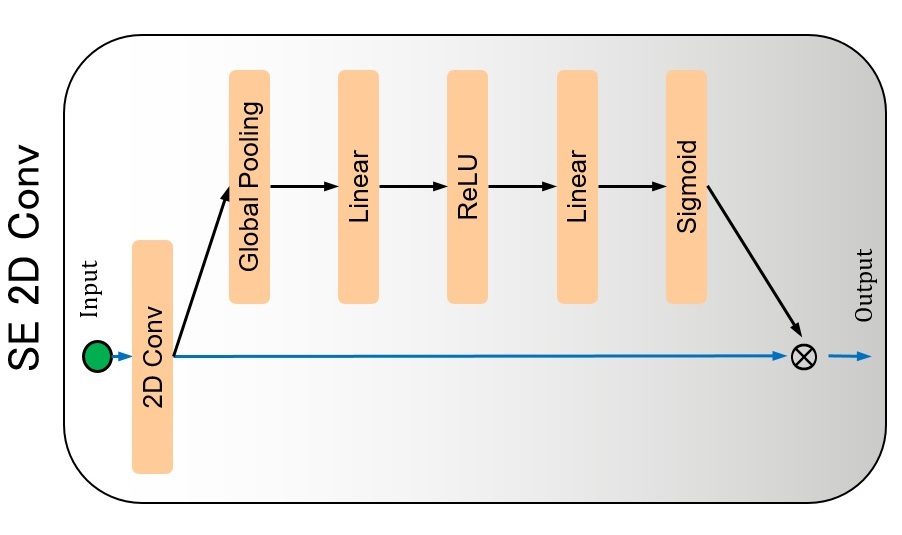}
\caption{SD Conv with squeeze-excitation mechanism (SE 2D Conv).}
\label{fig:selayer}
\end{figure}

A CRNN is used to map the Mel spectrogram to the frame-level and clip-level posterior probabilities of sound event presence. In order to improve the quality of representations by explicitly modeling the interdependencies between the channels of convolution layers in CRNN, the squeeze-excitation mechanism~\cite{hu2018squeeze} is introduced to CRNN for SED. The structure of 2D Conv with squeeze-excitation mechanism, which is called `SE 2D Conv' building block, is depicted in Fig.~\ref{fig:selayer}. The squeeze-excitation can learn to use global information to selectively emphasize informative features and suppress less useful ones. The feature maps are passed through a squeeze operation (Global Pooling) and an excitation operation (a simple gating mechanism with a Sigmoid activation and ReLU function using two fully connected Linear layers) to produces a collection of per-channel modulation weights, which are applied to the original feature maps to generate the output of the `SE 2D Conv' which can be fed directly into subsequent layers of the network.

%Further, in order to effectively capture long-range context dependencies in sound signal, we use gated linear units (GLUs) instead of commonly rectified linear units (ReLUs) or leaky ReLUs as nonlinear activations. Fig.~\ref{fig:gcnns} shows the structure of a GLU, where $i$ and $o$ are the input and output, $W$, $b$, $W_g$, and $b_g$ are learned parameters, $\sigma$ is the sigmoid function and $\otimes$ is the element-wise product between vectors or matrices.Similar to LSTMs, GLUs play the role of controlling the information passed on in the hierarchy.

%\begin{figure}[th]
%\centering
%\includegraphics[width=0.45\textwidth]{gcnns}
%\caption{Architecture of a GLU.}
%\label{fig:gcnns}
%
%\end{figure}

\begin{figure}[th]
    \centering
    \includegraphics[width=0.7\textwidth]{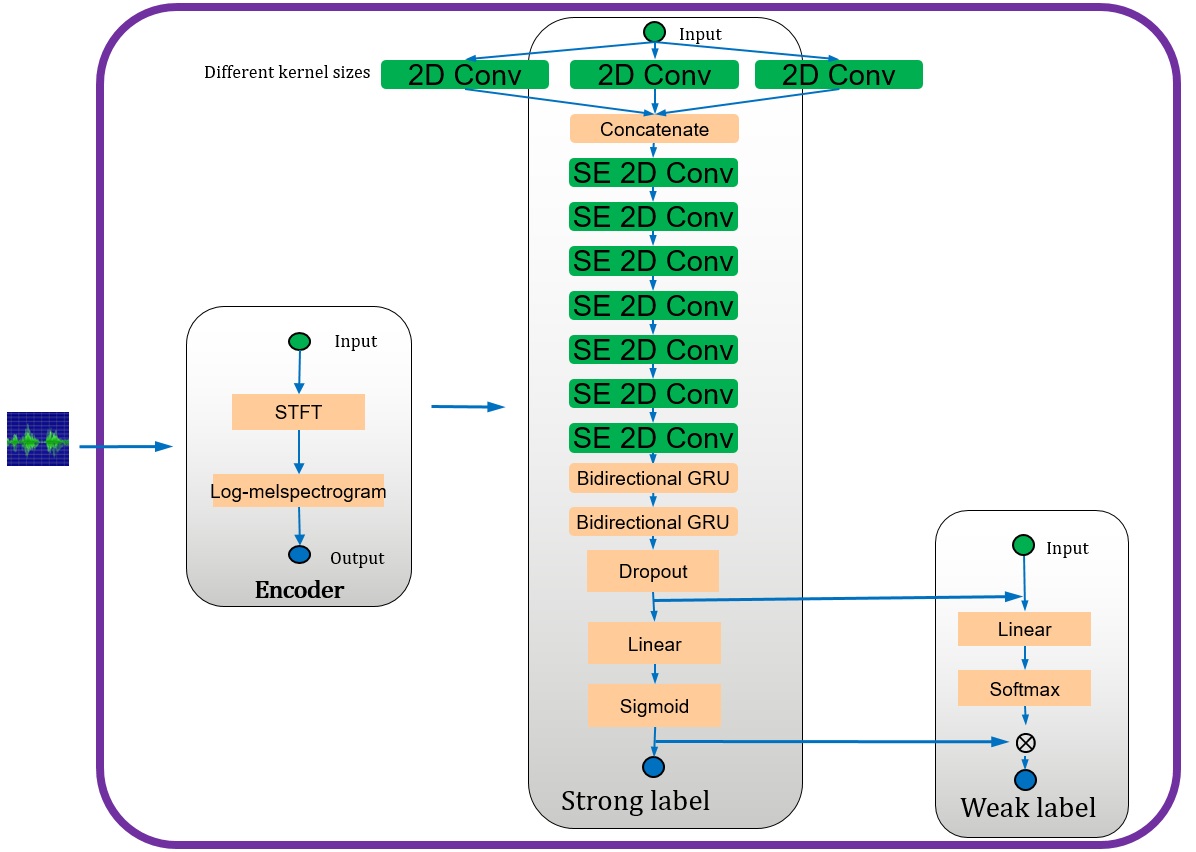}
    \caption{Architecture of the PSE-CRNN in Hodge and Podge.}
    \label{fig:psecrnn}
    \end{figure}

With the squeeze-excitation mechanism and GLUs, Fig~\ref{fig:psecrnn} presents the network architecture employed in our Hodge and Podge. In order to distinguish it from the traditional CRNN, here we call this backbone pyramid squeeze-excitation CRNN (PSE-CRNN). The audio signal is first converted to [128$\times$1024] log-Mel spectrogram to form the input to the network. Here in the Fig~\ref{fig:psecrnn}, the horizontal arrow from the `encoder' block to the `strong label' block actually means the output of the `encoder' block is the input to the `strong label' block. The first half of the `strong label' block network consists of one pyramid plain convolutional layer with three parallel 2D convolutional modules of different kernel sizes 3, 5, and 7 (considering different sound events have different spans in time-frequency domain), and seven squeeze-excitation gated convolutional layers, for which the kernel sizes are 3, the paddings are 1, the strides are 1, and the numbers of filters are [16, 32, 64, 128, 128, 128, 128] respectively, and the poolings are [(2, 2), (2, 2), (1, 2), (1, 2), (1, 2), (1, 2), (1, 2)] respectively. Pooling along the time axis is used in training with the clip-level and frame-level labels. The gated SE convolutional blocks are followed by two bidirectional gated recurrent units (GRU) layers containing 64 units in the forward and backward path, their outputs are concatenated and passed to the attention and classification layer which are described below.

As depicted in Fig.~\ref{fig:psecrnn}, the output of the bidirectional GRU layers is fed into both a frame-level classification block and an attention block (the `weak label' block in the figure) respectively. Thus there are two outputs in this PSE-CRNN. The output from bidirectional GRUs followed by dense layers with sigmoid activation is considered as the sound event detection result.  This output can be used to predict event activity probabilities. The other output from the `weak label' block is the weighted average of the element-wise multiplication of the attention, considering as audio tagging result. Thus the final prediction for the weak label of each class is determined by the weighted average of the element-wise multiplication of the attention and classification block output of each class.

\subsection{Composition Consistency Training}
\label{ssec:cct}

Under the framework of hybrid supervised sound event detection described in Section~\ref{sec:sed} and shown in Fig.~\ref{fig:mean_teacher}, we propose to introduce composition consistency training (CCT) to make full use of unlabeled clips. The intuition of CCT is that by weighted combining the original data and the augmented data, the prediction results of the model will still be the same weighted combination, whether it is on labeled data or unlabeled data. During training, there are 24 audio log-Mel spectrograms in each batch, of which 6 are weakly labeled, 6 are synthetic, and the remaining 12 are unlabeled. The model parameters $\theta$ are updated to encourage consistent predictions
\begin{equation}
f_\theta(\text{Mix}_\lambda(u_j,u_k))\approx \text{Mix}_\lambda(f_{\theta}(u_j),f_{\theta}(u_k)), \nonumber
\end{equation}
and correct predictions for labeled examples, where
\begin{equation}
\text{Mix}_\lambda(a,b)=\lambda a + (1-\lambda)b \nonumber
\end{equation}
is called the MixUp~\cite{zhang2017mixup} of two log-Mel spectrograms $u_j$ and $u_k$, and in each batch we sample a random $\lambda$ from Beta($\alpha$; $\alpha$) (e.g. $\alpha=1.0$ in all our settings). It should be noted that $\lambda$ is different for each batch. In CCT, we perform MixUp of sample pair and their corresponding labels (or pseudo labels predicted by the PSE-CRNNs) in both the supervised loss on labeled examples and the consistency loss on unsupervised examples. In each batch, the weakly labeled data, synthetic data, and unlabeled data are shuffled separately to form a new batch. Then we use the CCT principle to generate new augmented data and labels with the corresponding clips in the original and new batches.

Thus in CCT, for weakly labeled data, clip level classification CE loss on MixUp of original data batch (D) and shuffled augmented data batch aug\_D is computed as
\begin{eqnarray*}
L_{w,\text{CCT}}=\text{CE}(f^w_\theta(\text{Mix}_\lambda(\text{D},\text{aug\_D})), \\ \text{Mix}_\lambda(f^w_{\theta}(\text{D}),f^w_{\theta}(\text{aug\_D}))),
\end{eqnarray*}
where $\theta$ is the parameter of PSE-CRNN model and $f^w$ is the clip level posterior probability output. For synthetic strongly labeled data, frame level classification CE loss on MixUp data of original batch and shuffled aug\_D is computed as
\begin{eqnarray*}
L_{s,\text{CCT}}=\text{CE}(f^s_\theta(\text{Mix}_\lambda(\text{D},\text{aug\_D})), \\  \text{Mix}_\lambda(f^s_{\theta}(\text{D}),f^s_{\theta}(\text{aug\_D}))). \nonumber
\end{eqnarray*}
where $f^s$  is the frame level posterior probability output. For unlabeled data, clip level and frame level posterior probability consistency loss based between student model and teacher model on MixUp data of D and  shuffled aug\_D is calculated by mean square error  
\begin{eqnarray*}
L_{cw,\text{CCT}}=\|f^w_\theta(\text{Mix}_\lambda(\text{D},\text{aug\_D}))\\- \text{Mix}_\lambda(f^w_{\theta'}(\text{D}),f^w_{\theta'}(\text{aug\_D}))\|
\end{eqnarray*}
and
\begin{eqnarray*}
L_{cs,\text{CCT}}=\|f^s_\theta(\text{Mix}_\lambda(\text{D},\text{aug\_D}))\\- \text{Mix}_\lambda(f^s_{\theta'}(\text{D}),f^s_{\theta'}(\text{aug\_D}))\|
\end{eqnarray*}
respectively, here $\theta'$ is the parameter of teacher model.

The total loss
\begin{equation}
L_{CCT}=L_{w,\text{CCT}}+L_{s,\text{CCT}}+w(t)(L_{cw,\text{CCT}}+L_{cs,\text{CCT}}) \nonumber
\end{equation}
where generally the $w(t)$ changes overtime to make the consistency loss initially accounts for a very small proportion, and then the ratio slowly becomes higher. $w(t)$ has a maximum upper bound, that is, the proportion of consistency loss does not tend to be extremely large. With different maximum upper bound of consistency weight $w(t)$, the trained model has different performances.

\subsection{Multi-Hot MixMatch Training}
\label{ssec:mmt}

We also try to introduce the latest semi-supervised learning principles in MixMatch~\cite{berthelot2019mixmatch}. MixMatch introduces a single loss that unifies entropy minimization, consistency regularization, and generic regularization approaches to semi-supervised learning. Unfortunately, MixMatch can only be used for one-hot classification, not suitable for our situation, where there may be several events in a single audio clip. Thus we introduce multi-hot MixMatch (M$^3$), our proposed semi-supervised learning method to simplify MixMatch for sound event detection.

Original MixMatch uses a Sharpen operation to let the average predictions across all augmentations approaching a Dirac (“one-hot”) distribution. Our multi-hot MixMatch discarded this average and Sharpen operation. During training, 2 different augmentations, for example, denoted as aug\_A and aug\_B, are generated for each sample in the batch, and multi-hot MixMatch to do MixUp only between the augmentations of the same data type. That means we perform interpolation of sample pairs and their corresponding labels (or pseudo labels predicted by the PSE-CRNNs) in both the supervised loss on labeled examples and the consistency loss on unsupervised examples. In each training batch, the original weakly labeled data, synthetic strongly labeled data, and unlabeled data in aug\_A and aug\_B are shuffled separately to form a new batch. For original weakly labeled data, clip level classification
CE loss on MixUp data from original data batch (D) and shuffled aug\_A is computed as
\begin{equation}
L_{w,\text{M}^3}=\text{CE}(f^w_\theta(\text{Mix}_\lambda(\text{D},\text{aug\_A})),\\ \text{Mix}_\lambda(f^w_{\theta}(\text{D}),f^w_{\theta}(\text{aug\_A}))). \nonumber
\end{equation}
For synthetic strongly labeled data, frame level classification CE loss on MixUp data from original batch
and shuffled aug\_B is computed as
\begin{equation}
L_{s,\text{M}^3}=\text{CE}(f^s_\theta(\text{Mix}_\lambda(\text{D},\text{aug\_B})),\\ \text{Mix}_\lambda(f^s_{\theta}(\text{D}),f^s_{\theta}(\text{aug\_B}))). \nonumber
\end{equation}
For unlabeled data, clip level and frame level posterior probability consistency loss based between student model and teacher model on MixUp data from shuffled aug\_A and shuffled aug\_B is calculated by mean square error
\begin{eqnarray*}
L_{cw,\text{M}^3}=\|f^w_\theta(\text{Mix}_\lambda(\text{aug\_A},\text{aug\_B})) \\- \text{Mix}_\lambda(f^w_{\theta'}(\text{aug\_A}),f^w_{\theta'}(\text{aug\_B}))\| \nonumber
\end{eqnarray*}
and
\begin{eqnarray*}
L_{cs,\text{M}^3}=\|f^s_\theta(\text{Mix}_\lambda(\text{aug\_A},\text{aug\_B})) \\- \text{Mix}_\lambda(f^s_{\theta'}(\text{aug\_A}),f^s_{\theta'}(\text{aug\_B}))\|
\end{eqnarray*}
respectively, here $\theta'$ is the parameter of teacher model.

The total loss
\begin{equation}
L_{\text{M}^3}=L_{w,\text{M}^3}+L_{s,\text{M}^3}+w(t)(L_{cw,\text{M}^3}+L_{cs,\text{M}^3}), \nonumber
\end{equation}
where generally the $w(t)$ changes, which is consistent with the functions and settings of CCT in Section~\ref{ssec:cct}.

\subsection{Model Ensemble}
\label{ssec:ensemble}

To further improve the performance of the system, we use ensemble methods to fuse different models. The main differences between the single models are the maximum values of the consistency loss weight. An ensemble model is constructed by averaging the outputs of several different models with different maximum consistency coefficients.

\section{Experiments and Results}
\label{sec:results}

\subsection{Dataset}

We evaluated our Hodge and Podge on the  dataset of DCASE 2019 challenge task 4. The datasets are from AudioSet~\cite{gemmeke2017audio}, FSD~\cite{fonseca2017freesound} and SINS dataset~\cite{dekkers2017sins}. The aim of this task is to investigate whether real but weakly annotated data or synthetic data is sufficient for designing sound event detection systems. There are a total of 1578 real audio clips with weak labels, 2045 synthetic audio clips with strong labels, and 14412 unlabeled in domain audio clips in the development set, while the evaluation set contains 1168 audio clips. Audio recordings are 10 seconds in duration and consist of polyphonic sound events from 10 sound classes. For further information about that dataset, such as the number of items per class, distribution over classes, properties of sound events, and specific sound characteristics, please refer~\cite{turpault2019sound}. 

The evaluation metric for this task is based on the event-based F-score~\cite{mesaros2016metrics}. The predicted events are compared to a list of reference events by comparing the onset and offset of the predicted event to the overlapping reference event. The onset of the right predicted event should be within 200 ms collar of the onset of the reference event and its offset is within 200 ms or 20\% of the event length collar around the reference offset.

%\subsection{Evaluation Metric}

%The evaluation metric for this task is based on the event-based F-score~\cite{mesaros2016metrics}. The predicted events are compared to a list of reference events by comparing the onset and offset of the predicted event to the overlapping reference event. If the onset of the predicted event is within 200 ms collar of the onset of the reference event and its offset is within 200 ms or 20\% of the event length collar around the reference offset, then the predicted event is considered to be correctly detected, referred to as true positive. If a reference event has no matching predicted event, then it is considered a false negative. If the predicted event does not match any of the reference events, it is considered a false positive. In addition, if the system partially predicts an event without accurately detecting its onset and offset, it will be penalized twice as a false positive and a false negative. The following equation shows the calculation of the F-score for each class.
%\begin{equation}
%F_c= \frac{2TP_c}{2TP_c+FP_c+FN_c}, \nonumber
%\end{equation}
%where $F_c, TP_c, FP_c, FN_c$ are the F-score, true positives, false positives, false negatives of the class c respectively. The final evaluation metric is the average of the F-score for all the classes.

\subsection{Systems}

Four systems are evaluated and compared across the above conditions:
\begin{itemize}
\item
HODGEPODGE: The solution provided by~\cite{shi2019hodgepodge}, which ranked third in the task 4~\cite{dcase2019task4results}. An ensemble model is constructed by averaging the outputs of different models with different maximum consistency coefficients in the interpolation consistency training principle~\cite{verma2019interpolation}.
\item
\textbf{Hodge}: The method proposed in Section~\ref{sec:method} using PSE-CRNN with multi-hot MixMatch training principle~\ref{ssec:mmt}.
\item
\textbf{Podge}: The method proposed in Section~\ref{sec:method} using PSE-CRNN with composition consistency training principle~\ref{ssec:cct}.
\item
Baseline: The official solution provided by~\cite{turpault2019sound} is based on a `Mean Teacher'-type algorithm~\cite{jiakai2018mean} with plain CRNN, which is as described in Section~\ref{sec:sed}. 
\end{itemize}

\subsection{Results}

The systems are evaluated with macro-averaged event-based F-score~\cite{mesaros2016metrics}. Practically the predicted frame-level strong label result should be continuous, thus we get a smooth prediction result through a 1-dimensional median filter in the time dimension. The median window size in Table~\ref{tab:median_window} indicates the size, which gives the shape that is taken from the input original predictions, at every element position, to define the input to the filter function. Since the official test set is not public, we use the official validation set as the test set, and divide the training set into 9:1, as our own training set and validation set. In this experiment, HODGEPODGE~\cite{shi2019hodgepodge} is chosen as the baseline. Table~\ref{tab:median_window} shows the final macro-averaged event-based evaluation results on the official validation set compared to the baseline system. In the DCASE2019 task 4 challenge, on the official test set, HODGEPODGE achieved 42.0\% performance, while the first place was 42.7\%~\cite{dcase2019task4results}. Compare them in parallel, HODGEPODGE showed 0.7\% gap with the state-of-the-art on the official test set, while it can be seen from the Table~\ref{tab:median_window} our Hodge achieved an average performance improvement of 1.6\% compared to HODGEPODGE on the official validation set. This indicates that our Hodge has achieved state-of-the-art performance. It can be seen from the Table~\ref{tab:median_window}, the attempts made in Hodge and Podge do improve performance. Hodge and Podge are better than HODGEPODGE in all different median window sizes, and the average absolute improvement in F-score was about 1.6\% and 0.8\% respectively.

%Since HODGEPODGE showed smallest gap (0.5\%) between performances on development dataset and dataset~\cite{dcase2019task4results} among all other submissions, it is believed HODGEPODGE series methods have the least overfitting on this task, and the results on validation data are reliable that can be generalized to testing data. 

%\footnotetext[1]{Although we have requested, the DCASE committee is still discussing (at the submission moment of this paper) how to expose the test data in the official competition, so all the results listed here are on the validation set.}

\begin{table}
    \centering
    \begin{tabular}{|c|c|c|c|c|}
        \hline
    Median window size   & 7  & 9 & 11  & 13\\
    \hline
    HODGEPODGE & 41.1\%	& 41.7\%	&41.7 \%&41.4\%\\
 \hline
    \textbf{Hodge} & 43.0\%	 & 43.4\%	&43.2\% &42.8\%\\
 \hline
    \textbf{Podge} & 42.7\%	 & 42.4\%	&42.2\% &42.0\%\\
 \hline
    DCASE baseline & \multicolumn{4}{c}{25.8\%} \\
     \hline
    \end{tabular}
    \caption{The performance of Hodge and Podge on validation data set under 
    different median window size.}
    \label{tab:median_window}
    \end{table}

\section{Conclusions}
\label{sec:conc}

In this paper, we proposed a method called Hodge and Podge for sound event detection using only weakly labeled, synthetic and unlabeled data. Our approach is based on CRNN, whereby 
we introduce pyramid squeeze-excitation mechanism, composition consistency training, and multi-hot MixMatch consistency training with temporal-frequency augmentation to leverage the information in audio data that are not accurately labeled. The final F-score of our system on the official validation set is 43.4\%, which is significantly higher than the score of the baseline system which is 25.8\%.

\bibliographystyle{splncs03}
\bibliography{sed}

\end{document}